\documentclass[prl,superscriptaddress,showpacs,twocolumn]{revtex4}
\usepackage{graphicx}
\usepackage{dcolumn}
\usepackage{amsmath}
\usepackage{amssymb}
\usepackage{color}

\begin{document}

\title{Kondo resonance for orbitally degenerate systems}
\author{A. K. Zhuravlev}
\affiliation{Institute of Metal Physics, 620219 Ekaterinburg,
Russia} \affiliation{University of Nijmegen, NL 6525 ED Nijmegen,
The Netherlands}
\author{V. Yu. Irkhin}
\affiliation{Institute of Metal Physics, 620219 Ekaterinburg,
Russia}\affiliation{Department of Physics, Uppsala University,
Box 530, SE-751 21 Uppsala, Sweden}
\author{M. I. Katsnelson}
\affiliation{Department of Physics, Uppsala University, Box 530,
SE-751 21 Uppsala, Sweden} \affiliation{University of Nijmegen,
NL 6525 ED Nijmegen, The Netherlands}
\author{A. I. Lichtenstein}
\affiliation{University of Nijmegen, NL 6525 ED Nijmegen, The
Netherlands} \affiliation{Institute of Theoretical Physics,
University of Hamburg, Jungiusstrasse 9, 20355 Hamburg, Germany}

\begin{abstract}
Formation of the Kondo state in general two-band Anderson model
has been investigated within the numerical renormalization group
(NRG) calculations. The Abrikosov-Suhl resonance is essentially
asymmetric for the model with one electron per impurity (quarter
filling case) in contrast with the one-band case. An external
magnetic (pseudo-magnetic) field breaking spin (orbital)
degeneracy leads to asymmetric splitting and essential broadening
of the many-body resonance. Unlike the standard Anderson model,
the ``spin up'' Kondo peak is pinned against the Fermi level, but
not suppressed by magnetic field.
\end{abstract}


\maketitle

The Kondo problem is one of most fascinating and important issues
in condensed matter theory. It was originally formulated for
explanation of the resistivity minimum in metallic alloys
\cite{kondo} due to the scattering of conduction electrons by a
magnetic impurity, and later generalized to various cases. The
Kondo effect turns out to be a key phenomenon of the heavy fermion
behavior \cite{hewson}, anomalous electronic properties of
metallic glasses at low temperatures \cite{cox}, quantum dots
\cite{qd}, and many other correlated electron problems. One of the
main results of this theory is the formation of a resonance with
small energy scale (so-called, the Kondo temperature $T_K$) near
the Fermi energy due to the scattering of conduction electrons by
local quantum systems with internal degrees of freedom. Originally
this resonance (usually called the Kondo, or Abrikosov-Suhl
resonance; for a review, see Ref. \onlinecite{hewson}) could be
experimentally investigated only indirectly, through the
temperature dependence of thermodynamic and transport properties
of metals. However, novel tunneling spectroscopy, in particular scanning tunneling
microscopy (STM), is now able to directly visualize the Kondo resonance
\cite{STM,coral,kol}. At the same
time,  theoretical investigation of the spectral density for Kondo
systems is a much more complicated problem than the calculation of
thermodynamic properties. Indeed, in the latter case exact
analytical results can be derived by the Bethe Ansatz \cite{wieg},
whereas the electron density of states is considered mainly by
some approximate methods or numerically, see, e.g., Refs.
\onlinecite{GS,largeN,kasuya,costi1,kroha,costi2,logan0,logan}.
The splitting of the Kondo resonance by external magnetic field
was investigated by Bethe Ansatz for the $s$-$d$ exchange model
\cite{moore}. These results can be used to verify different
approximate schemes demonstrating some difficulties, for example, with
the well-known non-crossing approximation (NCA) \cite{largeN}.

Generally speaking, we have now a complete and satisfactory theory
of the Kondo resonance for the prototype case of purely spin
scattering. On the other hand, the information about the systems
with orbital degrees of freedom is still insufficient. A so-called
``orbital Kondo resonance'' has been considered theoretically, for
atomic two-level systems in metallic glasses, for quadrupolar
degrees of freedom in some uranium-based compounds \cite{cox}, for
high-temperature superconductors \cite{IKT}, and for double
quantum dot systems \cite{double}. Recently the phase diagram of
the Anderson model with orbital degrees of freedom has been
investigated by the numerical renormalization group (NRG) method
\cite{fabrizio}. Spin (pseudospin) susceptibility for the double
quantum dot model has been investigated by this technique in Ref.
\onlinecite{double}. However, electron spectral density has not
been calculated. It is worthwhile to note that investigation of
dynamical properties for effective impurity models is of the
crucial importance for the dynamical mean-field theory (DMFT)
approach, in particular, to describe the metal-insulator
transition and related phenomena \cite{dmft}. For one-band case
the NRG method was applied to DMFT problem in Ref.
\onlinecite{bulla}.

Direct observation of the orbital Kondo resonance on Cr(001)
surface by the STM measurements \cite{kol}, as well as a relevance
for multiple quantum dot systems \cite{double}, makes the issue
about the shape of the Kondo resonance and about effects of
``pseudomagnetic'' field which breaks the orbital degeneracy
especially actual. Here we investigate the problem of orbital
Kondo resonance by the NRG approach \cite{Wilson}. We shall
demonstrate that in the ``orbital'' Kondo case the resonance has
an essential asymmetry with respect to the Fermi energy, in
qualitative agreement with the experimental observations
\cite{kol}.

We start from the two-band Anderson impurity model with the spin and
orbital rotationally invariant Hamiltonian \cite{dworin}
\begin{equation}
H=\sum_{\mathbf{k}a\sigma }\left[ \varepsilon _{\mathbf{k}}c_{\mathbf{k}%
a\sigma }^{\dagger }c_{\mathbf{k}a\sigma }+V\Big(f_{a\sigma }^{\dagger }c_{%
\mathbf{k}a\sigma }+c_{\mathbf{k}a\sigma }^{\dagger }f_{a\sigma }\Big)\right]
+H_\mathrm{imp}  \label{eq:siam}
\end{equation}
where
\begin{eqnarray}
H_\mathrm{imp}=\sum_{a\sigma }\Big(\varepsilon _{\mathrm{f}}-\frac{h\sigma}{2}\Big) %
f_{a\sigma }^{\dagger }f_{a\sigma }+\frac{U+J}2\sum_{a\sigma }n_{a\sigma
}n_{a-\sigma } +  \notag \\
\sum_{a\neq b,\sigma }\left( \frac U2n_{a\sigma }n_{b-\sigma }+\frac{U-J}%
2n_{a\sigma }n_{b\sigma }-\frac J2f_{a\sigma }^{\dagger }f_{a-\sigma
}f_{b-\sigma }^{\dagger }f_{b\sigma }\right)  \notag
\end{eqnarray}
Here $a,b$ = 1,2 and $\sigma= \uparrow ,\downarrow$ are orbital and spin
indices, correspondingly, $c_{\mathbf{k}a\sigma }^{\dagger }$($c_{\mathbf{k}%
a\sigma }$) denote the creation (annihilation) operators for
$a$-orbital states with spin $\sigma $ and energy $\varepsilon
_{\mathbf{k}}$ (we take the rectangle band with half
width $D=2$), $f_{a\sigma }^{\dagger }$($f_{a\sigma }$) those
for impurity states of $a$-orbital with spin $\sigma $ and energy
$\varepsilon _{\mathrm{f}}$, $n_{a\sigma }=f_{a\sigma  }^{\dagger }f_{a\sigma
}$, $h$ is magnetic field. Since the spin and orbital degrees of
freedom are symmetric in the Hamiltonian, $h$ may be a
pseudomagnetic field (e.g., for the orbital Kondo effect on
Cr(001) surface where the potential of the atomic step edge breaks
the exact degeneracy between  $d_{xz}$ and $d_{yz}$ states
\cite{kol}). The Coulomb interaction and exchange parameter at the
impurity site are $U$ and $J$, and two electron subsystems are
coupled via a hybridization parameter $V$. Note that in solids
the orbital moment conservation is not an exact property
and therefore some additional terms in the Hamiltonian may appear
\cite{blandin}. However, we omit them since the problem turns out
to be numerically very cumbersome even for the rotationally
invariant Hamiltonian.

To calculate the spectral properties of impurity we use the NRG
technique which is described in details in Refs.
\onlinecite{Wilson,hewson,kasuya,costi}. Here we emphasize the new
aspects for multi-orbital Anderson model.

As usually, we start from the solution of the isolated impurity
problem. As an initial step of the RG procedure, we add the first
conduction electron site, diagonalize the Hamiltonian matrix for
this Hilbert space, and thus obtain new eigenstates. Then such a
procedure should be repeated until reaching a fixed point.
Dimension of the Hilbert space within one NRG-iteration increases
by factor 16 instead of 4 for the one-band case. By using an
appropriate symmetry of the problem we were able to find the whole
spectrum for the Hilbert space with the dimensionality of 48000
states, which gives a possibility to keep at each step about 3000
states
\cite{method}.

Due to the NRG discretization scheme, the electron spectral function
\begin{equation*}
\rho _{f}(\omega )=-\frac{1}{\pi }\mathrm{Im}G(\omega +i0)\ ,\ \ \ \ \ \
G(z)=\langle \langle f_{a\sigma }|f_{a\sigma }^{\dagger }\rangle \rangle _{z}
\end{equation*}
is given by a set of $\delta $-functional peaks at the frequencies $\omega
_{n}$. Standard NRG practice consists of the Gaussian broadening of the
spectral function on a logarithmic scale \cite{kasuya}. Since the point $%
\omega =0$ plays  special role in such a scheme, we used more conventional
Gaussian broadening

\begin{equation}
\delta (\omega -\omega _{n})\rightarrow \frac{1}{b_{L}\sqrt{\pi }}\exp \left[
-\frac{(\omega -\omega _{n})^{2}}{b_{L}^{2}}\right] ,
\end{equation}
smearing being changed depending on the iteration number $L$, namely, $%
b_{L+2}=b_{L}/\Lambda $ (where the NRG cutoff parameter $\Lambda =2$ have
been used).

\begin{figure}[htbp]
\includegraphics[
width=3.0in,angle=0]{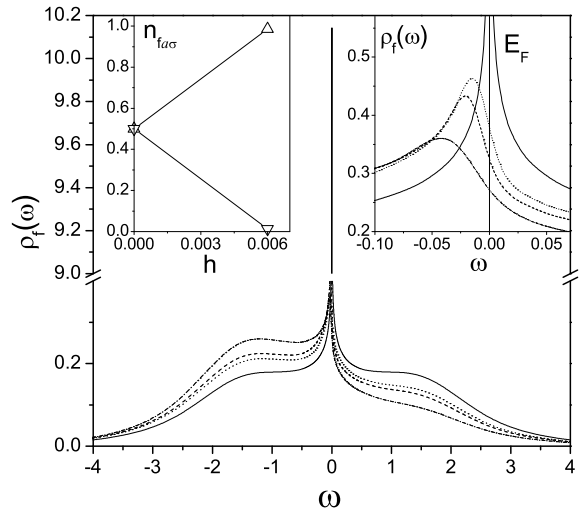} \caption{The density of states for
the half-filled case, $\varepsilon _{\mathrm{f}}=-3,V=0.2,U=2,J=0,
h=0$ (solid) and symmetric splitting (only the spin-up DOS) in the
magnetic field $h=0.006$ (dot), $h=0.01$ (dash) and $h=0.02$
(dash-dot). Occupation number per orbital and spin are shown in
the left insert (total $n_{f}=2$). Density of states at the Fermi
level according to the Friedel sum rule (\ref{rho}) is $\protect\rho
_{f}\left( 0\right) =10.13$. Right insert shows the DOS in the vicinity
of the Fermi level.} \label{Fig1}

\end{figure}

The Kondo regime corresponds to the case where $|\varepsilon _{\mathrm{f}%
}|\gtrsim 2\Gamma $ \cite{costi1}, $\Gamma =\pi V^{2}\rho $ being
the one-particle resonance width for the localized electrons,
$\rho $ is the bare DOS of conduction electrons at $E_{F}$. The
density of states for the half-filled case ($n_f=2$) and for almost
quarter-filled case ($n_f\approx 1$) are shown in Figs. 1 and 2, respectively.
Although the Hund interaction $J$ was estimated to have
considerable value \cite{kol}, this quantity is irrelevant for the
quarter-filled case; our RG calculations confirmed that $J$ does
not influence essentially the results (see insert in Fig. 2). The
main difference of the NRG results obtained here with those for
the nondegenerate Anderson model is that the Kondo peak 
is not centered at the Fermi energy $E_{F}=0$. An explanation of
this deviation is given by the Friedel sum rule for the phase
shifts $\eta _{l}$: $2(2l+1)\eta _{l}/\pi =n_{f}$ \cite{friedel},
$2l+1$ being the number of orbital channels and $n_{f}$ the total
number of localized electrons. It is important that due to locality of the
self-energy the value of $\rho _{f}(0)$ does not change in comparison with
the non-interacting model ($U=0$) and is equal to
\begin{equation}
\rho _{f}(0)=\frac{1}{\pi \Gamma }\sin ^{2}\left( \frac{\pi n_{f}}{N}\right)
,  \label{rho}
\end{equation}
where $N$ is the degeneracy factor \cite{zawad}. For the standard
$SU(2)$ Kondo model for $S=1/2$, as well as for a degenerate
half-filled model, the phase shift at $E_{F}$ is close to $\pi /2$
which means the strongest possible scattering at the top of the
resonance peak. In this case the asymmetry of the Kondo resonance
with respect to $E_{F}$ should be very weak even for the
nonsymmetric Anderson model, in agreement with recent
computational results \cite{logan}, and with our Fig. 1. Thus the
large value of electronic effective mass and linear specific heat
is owing to renormalization of the residue of the electron Green's
function,
\begin{equation}
Z=\left( \left. 1-\frac{\partial \mathrm{Re}\Sigma (E)}{\partial E}\right|
_{E=E_{F}}\right) ^{-1}  \label{z}
\end{equation}
On the other hand, for $N>2$ and $n_{f}\approx 1$ the top of the
f-electron peak shifts above $E_{F}$ \cite{hewson}. In particular,
in the $SU(N)$ Anderson model (or in the equivalent
Coqublin-Schrieffer model) with infinitely large $N$
\cite{GS,largeN} the high peak lies \textit{completely} above the
Fermi level, being shifted by the value of the order of $T_{K}$
(note that in the limit $N\rightarrow \infty $ \ this is just a
delta-like peak). NRG calculations for $SU(N)$ Anderson model
\cite{kasuya} give qualitatively similar results. However, the
values of $\rho _{f}(0)$ calculated in Ref. \cite{kasuya} turn out
to be by about 30$\%$ smaller than given by Eq.(\ref {rho}),
probably, due to insufficient accuracy. Our results for
double-degenerate model give $\rho _{f}(0)$ in the perfect
agreement with the Friedel sum rule (see captions to Fig. 1).

\begin{figure}[htbp]
\includegraphics[
width=3.0in,angle=0]{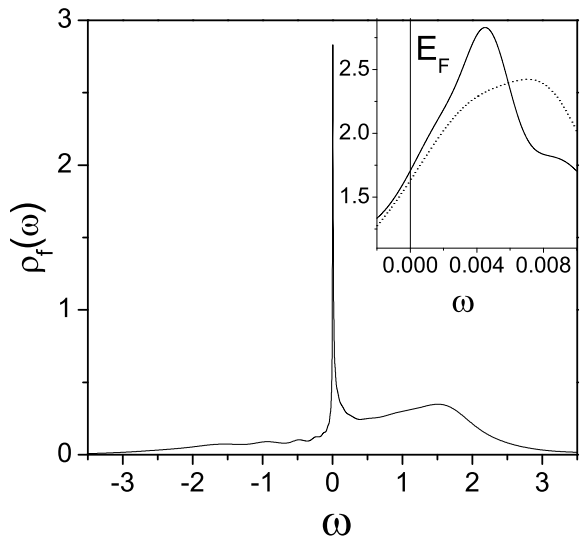}
\caption{The density of states for the almost quarter-filled case for $%
\protect\varepsilon_{\mathrm{f}}=-1,U=2,J=0,V=0.35$. Occupation
number per
orbital and spin is 0.258 (total $n_f=1.032$).
Insert shows the DOS in the vicinity of the Fermi level;
the dash line corresponds to the case of non-zero exchange parameter
$J=0.2$.} \label{Fig2}
\end{figure}

The NRG calculations with the external magnetic field $h$ were
also performed. Since we are interested only in a small energy
region, shortcomings of the NRG scheme, which occur in the
presence of magnetic field at large excitation energies $\omega $
\cite{hofst}, are not important here. There are several
qualitatively different regimes of magnetic splitting in the
double-degenerate Anderson model:

(i) a half-filled case where $n_f=2$ due to electron-hole symmetry
even in the presence of external magnetic field. The usual
symmetrical splitting takes place, the Kondo peaks becomes low and
broad (Fig. 1), similar to the behavior in the non-degenerate
symmetric Anderson model \cite{logan}.

(ii) a nearly quarter-filled case: $n_f\approx 1$. The splitting
of the Kondo peak is asymmetric. One can see from Fig. 3 that the
upper ``spin down'' peak becomes low and broad, as well as in the
symmetric case. At the same time, the lower ``spin up'' peak is
not suppressed (as in the standard Kondo model), but tends to the
Fermi level and becomes more high (however, the situation changes 
with decreasing $V$: the height and area of spin-up 
peak decreases 
considerably in strong magnetic field). The density of states at
$E_F$ remains high up to very strong fields, so that the partial
f-occupation numbers depend strongly on $h$. For very strong fields
the peak corresponding to spin down states is completely
suppressed ($n_{fa \downarrow} \rightarrow 0$), and only orbital
Kondo resonance between spin up states survives since 
$n_{f a\uparrow} \rightarrow 1/2$, in contrast with the case (i) 
where $n_{f a\uparrow} \rightarrow 1$.

(iii) the intermediate valence regime ($n_{f}$ is essentially
non-integer). Instead of the three-peak structure characteristic
for the Kondo regime, we have one peak which is split in magnetic
field (Fig. 4).

\begin{figure}[h]
\includegraphics[
width=3.in,angle=0]{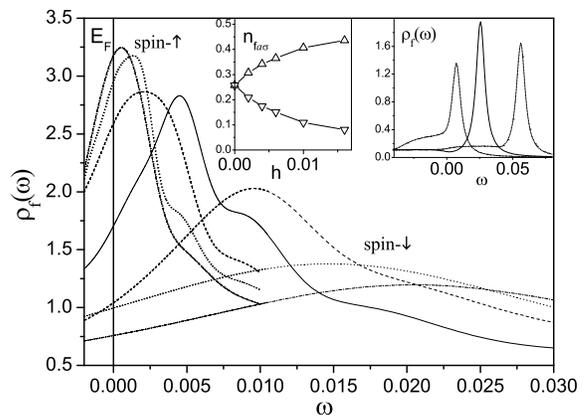} \caption{The effect of the magnetic field on
the density of states for the same model parameters as in Fig. 2.
The magnetic field values are $h=0$ (solid), 0.006 (dash), 0.01
(dot), 0.016 (dash-dot). The left insert displays the dependence
of occupation numbers on the magnetic field; the right one
shows the result of the approximation (\ref{sigh}) for
$\varepsilon_\mathrm{f}=-1, U=\infty, J=0, V=0.56, h=0$ (solid) and
$h=0.04$ (dot), the
smearing of the logarithm with $\delta = 0.004$ being introduced.
The value of $V$ is increased in comparison with the finite-$U$
case to obtain a comparable value of the Kondo temperature.}
\label{Fig3}
\end{figure}

\begin{figure}[h]
\includegraphics[
width=3.in,angle=0]{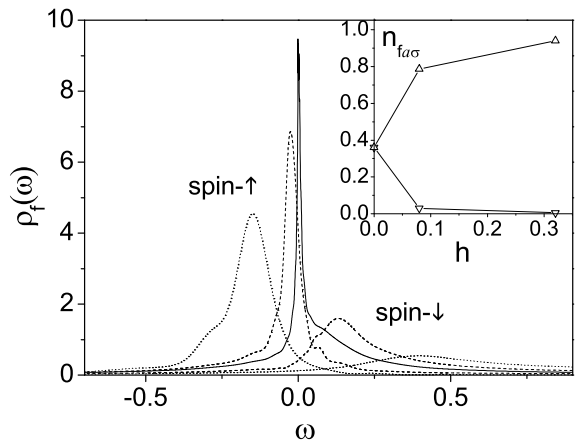}
\caption{The effect of the magnetic field on the density of states for $%
\protect\varepsilon_{\mathrm{f}}=-2,U=2,J=0,V=0.2$. The magnetic
field values are $h=0$ (solid), 0.08 (dash), 0.32 (dot). The
insert shows the spin-occupation number as function of magnetic
fields.} \label{Fig4}
\end{figure}

The NRG method can provide a detailed information about the Kondo
resonance in the two-band Anderson model. For larger number of
orbitals, numerical calculations become impractical for a reliable
treatment of the Kondo problem. To understand qualitatively the
behavior of the multiband Kondo system in magnetic field, we
consider a simple degenerate Anderson model with $U=\infty $ where
the peak lies also above the Fermi level \cite{largeN}. The
Hamiltonian reads
\begin{equation*}
\mathcal{H}=\sum_{\mathbf{k}m}t_{\mathbf{k}}c_{\mathbf{k}m}^{\dagger }c_{%
\mathbf{k}m}+\sum_{m}(\varepsilon _{\mathrm{f}}-h_{m})f_{m}^{\dagger
}f_{m}+V\sum_{\mathbf{k}m}\left[ c_{\mathbf{k}m}^{\dagger }f_{m}+h.c.\right]
\end{equation*}
where $f_{m}^{\dagger }=|m\rangle \langle 0|$ are the Hubbard's $X$%
-operators; the model (\ref{eq:siam}) corresponds to $m=a\sigma$
and $h_{a\uparrow}=h/2$, $h_{a\downarrow}=-h/2$. Using the
second-order perturbation theory for $X$-operators \cite{IKZ} one
can obtain (cf. also Refs. \cite{GS,largeN})
\begin{equation}
\begin{split}
\langle \!\langle f_{m}|f_{m}^{\dagger }\rangle \!\rangle _{E}=\langle
n_{0}+n_{m}\rangle \big[ E-\varepsilon _{\mathrm{f}}+h_{m} \\
-\rho
V^{2}\sum_{m^{\prime }\neq m}\ln \frac{D}{E+h_{m}-h_{m^{\prime }}}\big]
^{-1}  \label{sigh}
\end{split}
\end{equation}
This very simple approximation gives reasonable
\textit{qualitative} agreement with the numerically accurate NRG
results (see right insert in Fig. 3). It can be used for
interpretations of the computational results. For example,
according to Eq.(\ref{sigh}), the spin up peak does not intersect
the Fermi level with increasing magnetic field (at least, at not
too small $V$). Indeed, the logarithmic divergence of the
self-energy exists in our orbitally-degenerate model, the
contribution from the transitions between the degenerate states
being not cut at $h$.

To conclude, we have considered peculiarities of the Kondo
resonance in the orbital-degenerate case by a numerical
renormalization group technique. A possibility to calculate the
spectral properties for degenerate Anderson impurity model is
demonstrated, which gives a chance to extend the applicability
region of the NRG scheme in the DMFT approach beyond the one-band
case. 
Our version of the NRG can describe accurately the case
where the Kondo peak is shifted from the Fermi energy, which is a
generic case of multiband impurity model. The new features of the
orbital-degenerate model are related with a fact, that the Kondo 
resonance is not suppressed by external magnetic (or pseudomagnetic) field,
its splitting being essentially asymmetric.

The work was supported by the Netherlands Organization for
Scientific Research (NWO project 047.016.005) and the Russian Academy of 
Sciences, grant No.NSh-747.2003.2.


\begin{thebibliography}{99}
\bibitem{method}  Details of multiband NRG calculations will be published
elsewhere.
\end{thebibliography}
\end{document}